\documentclass[twocolumn,twoside,slac_two]{revtex4}
\usepackage{graphicx}
\usepackage{fancyhdr}
\def\ltsima{$\; \buildrel < \over \sim \;$}
\def\simlt{\lower.5ex\hbox{\ltsima}} 
\def\gtsima{$\; \buildrel > \over \sim \;$}
\def\simgt{\lower.5ex\hbox{\gtsima}} 

\begin{document}

\title{Large-Scale Radio and X-ray Jets in the Highest Redshift Quasars}

\author{C. C. Cheung\footnote{Jansky Postdoctoral Fellow}}
\affiliation{National Radio Astronomy Observatory, and Kavli Institute for 
Astrophysics \& Space Research, \\Massachusetts Institute of Technology, 
Cambridge, MA 02139, USA} 
\author{J. F. C. Wardle, N. P. Lee}
\affiliation{Physics Department, Brandeis University, Waltham, MA 02454, USA}

\begin{abstract}

We describe our program to search for and study the kilo-parsec scale
radio jets in a sample of high-redshift (greater than 3.4), flat
spectrum quasars 
using new and archival VLA data.  Two of these radio jets have been imaged with Chandra, 
and have X-ray counterparts and are briefly discussed.
These high-redshift sources are important
targets for testing current X-ray jet emission models for kpc-scale jets and follow-up
multi-wavelength observations will shed light on this problem.

\end{abstract}

\maketitle

\thispagestyle{fancy}


\section{Kiloparsec-scale X-ray Jets in AGN}

Chandra X-ray Observatory and Hubble Space Telescope observations have
established that X-ray and optical emission from kpc-scale radio jets in
active galactic nuclei (AGN) are common (e.g. Stawarz 2003).  In quasars,
the X-rays are widely interpreted as inverse Compton (IC) scattered
emission off the CMB by electrons in the jet emitting synchrotron
radiation at very low radio frequencies (Tavecchio et al. 2000;
Celotti et al. 2001). If this is the dominant X-ray production mechanism,
it implies quasar jets have large bulk velocities (Lorentz factors,
$\Gamma$$\sim$3--15) on the observed 10's -- 100's kpc-scale, so that the
electrons in the jet frame see a sufficiently boosted source of seed
photons to explain the X-rays. 

A natural consequence of the IC/CMB model is that high-z quasars should
have prominent X-ray jets (Schwartz 2002a) because of the strong
dependence of the CMB energy density on redshift: $f_{X}/f_{r}\propto$ U$_{\rm
CMB}$$\propto$(1+z)$^{4}$.  The recent detection of a bright X-ray jet in
the z=4.3 quasar GB~1508+5714 (Yuan et al. 2003, Siemiginowska et al. 
2003), with only a faint radio counterpart (\S~2), lends
support to this scenario.  This jet sees 1--2 orders of magnitude times
greater energy density from the CMB than jets at lower redshift
(z$\simlt$2), so its extreme redshift may account, to first order, for its
large X-ray to radio luminosity ratio (Cheung 2004). In the
framework of the IC/CMB model, variations in $B$ and jet Doppler factor,
$\delta$, can cause significant spread in the observed $f_{X}/f_{r}$ (both
along a given jet, and source-to-source), and would smear the
possible relation with redshift if only a limited redshift-range is studied. 

It is important to identify more jets in high-redshift quasars to observe
with both Chandra and the VLA in order to compare with lower-redshift
examples. Observations of these systems could help to distinguish between
the different models proposed to account for the X-ray emission -- one
would not, for instance, expect a z-dependence in the synchrotron X-ray jets.

Unfortunately, very little is known about the extended radio structure of
high-redshift quasars. Recent Chandra imaging of a number of z$>$4 radio
loud quasars do not reveal significant extended X-ray emission (Bassett et
al. 2004; Lopez et al. 2004), although there is no pre-existing information
on possible radio structures in these samples. 

We recently began a program to search for and study kpc-scale radio jets in
high-redshift quasars.  The overall goal of our work is to compile a
comparison sample to the X-ray jet detections at lower-redshift to test for a
bulk redshift dependence of $f_{X}/f_{r}$.  We do not aim for our sample
to be a complete one since the lower-z detections are inhomogenous -- we
require only that our high-z sample be comparable in radio luminosity and
that the jets are beamed (i.e. are flat-spectrum core-dominated quasars). 
Ongoing work from this program is discussed. 

\begin{figure*}[t]
\centering
\includegraphics[width=72mm,angle=-90]{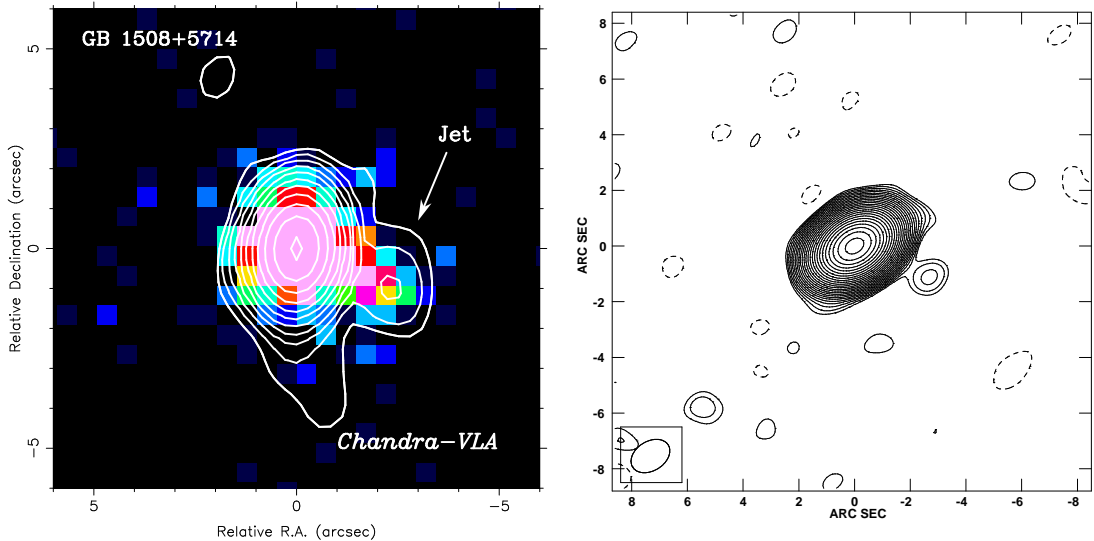}
\caption{[LEFT] Chandra X-ray image of GB~1508+5714 (color) with VLA 1.4
GHz contours overlaid (taken from Cheung 2004).
[RIGHT] New VLA 4.9 GHz map confirming the faint radio/X-ray
feature $\sim$2.5'' west of the core (see left panel). The lowest
contour level is 67 $\mu$Jy/bm (2$\sigma$), and subsequent ones increase by
factors of $\sqrt{2}$
up to a peak of 0.323 Jy per beam (1.50''$\times$1.05'' at PA=--57.8 deg).
}
\label{f1}
\end{figure*}

\section{\label{gb1508} The X-ray / Radio ``Jet'' in the z=4.3 Quasar 
GB~1508+5714}

The Chandra detection of an extended X-ray feature $\sim$2.5'' from the
z=4.3 quasar GB~1508+5714 (Hook et al. 1995) was reported independently by
two groups:  Siemiginowska et al. (2003) and Yuan et al. (2003).
Subsequently, a faint radio counterpart (1.2 mJy at 1.4 GHz) to the extended
X-ray feature was found (Cheung 2004), supporting the previous authors'
interpretation of the emission as from a jet associated with the quasar.
This makes GB~1508+5714 the most distant quasar with a {\it
kiloparsec-scale} jet detected at any wavelength\footnote{A similar single
jet feature has been disovered recently on smaller (VLBI) scales in the more
distant z=5.47 blazar Q0906+6930 by Romani et al. (2004).}. 

The jet knot in GB~1508+5714 shows an extremely high monochromatic X-ray
to radio luminosity ratio ($f_x$/$f_r$=158). This is large in comparison
to X-ray jets detected in lower redshift, z$\simlt$2 core-dominated quasars (range of 
$\sim$0.2--50). 
Since this observed ratio is strongly dependent on redshift in the IC/CMB model (e.g.
Schwartz 2002a), it suggests that this is an important and presumably a
dominant X-ray production mechanism in kiloparsec-scale quasar jets (see
Cheung 2004).

\subsection{\label{radiospec} Radio Spectrum of the Jet}

We have since obtained exploratory time VLA B-configuration observations of GB~1508+5714
at 4.9 and 8.4 GHz (program AC728).  A
total of $\sim$30 minutes integration per frequency was obtained, and the
measured off-source rms in the self-calibrated images are within 10$\%$ of
the theoretical noise level of 0.03 mJy/bm. We measured an inverted radio
spectrum ($\alpha$=--0.5 $\pm$0.1;  $F_{\nu}\propto\nu^{-\alpha}$) for the
radio core between 4.9--8.4 GHz at our observation epoch (January 2,
2004). 

The extended $\sim$2.5'' radio feature is confirmed at 5 GHz, where we
found a very faint counterpart (0.2 $\pm$0.04 mJy, a 5$\sigma$ detection;
Figure~1). With the previous 1.4 GHz detection, the 1.4--5 GHz radio
spectral index is 1.4$\pm$0.2, which is very steep in comparison to other
radio jet features (e.g. Bridle \& Perley 1984). Extrapolating this
spectrum to 8.4 GHz, the expected flux density is $\sim$0.1 mJy,
consistent with our previous limit at this frequency from archival data
(Cheung 2004), and non-detection in the new observation (5$\sigma$=0.2 mJy
at the position of the jet).  The steep observed radio spectrum is
probably due to the fact that we are sampling both higher rest-frame
energies (up to 44 GHz for the 8.4 GHz limit) and the increased efficiency of
radiative losses due to inverse Compton scattering on the cosmic microwave
background at this high-redshift. 

\subsection{\label{construct} Reconstructing the Electron Energy Spectrum in a 
One-zone Model}

Our new measurement of the jet radio spectrum allow us to rule out a
single power-law extrapolation of the radio data into the X-ray band,
since $\alpha_{\rm r} > \alpha_{\rm rx}$ (=0.73), as would be expected in
canonical synchrotron models.  If we assume the radio and X-ray emission
are produced co-spatially, we can apply a "one-zone" synchrotron (radio)
and IC/CMB (X-ray) model (Tavecchio et al.  2000) as done previously for
this jet ($\S$\ref{gb1508}).  Assuming also equipartition between the
relativistic particles and magnetic field, we can solve for the Doppler
factor, $\delta\sim$4, and magnetic field, $B\sim$25 $\mu$G assuming
$\alpha$=0.9, the measured X-ray spectral index (Cheung 2004). 

This interpretation then demands that we (or allows us to, depending on
one's vantage point) reconstruct the underlying electron energy spectrum
in some detail as follows:

\noindent $\bullet$ The radio spectrum is steeper than the X-ray one
($\alpha_{\rm x}\sim$0.9$\pm$0.3; Siemiginowska et al. 2003, Yuan et al.
2003) implying a break in the electron energy spectrum of
$\sim$0.5$\pm$0.4. 

\noindent $\bullet$ The detected IC/CMB X-ray emission is produced by
electrons with energy $\gamma\sim$ 100-1000 (0.3--5 keV) which emit 
synchrotron radiation at (unobservable) radio frequencies of 40 MHz and 
below. 

\noindent $\bullet$ The X-ray spectral index of 0.9 reveals the 
intrinsic
index of the relativistic electron population before the spectrum drops
steeply in the cm-band ($\gamma\sim$ 6$\times$10$^{3}$ to 10$^{4}$ for 1.4
and 5 GHz). 

The overall picture is that electrons with energies $\gamma\sim$ 100 to at least $10^{4}$
are present, and a break in the spectrum occurs between $\gamma\sim$1000--6000. This has
important implications for prospects of future multi-wavelength studies of kpc-scale jets: 

\noindent $\bullet$ The lowest energy electrons Comptonize CMB photons
into the optical band ($\gamma\sim$10). These electrons dominate the total
energy density in the jet and deep HST observations offer a possible probe
of their emission which would otherwise be unobservable at sub-MHz
frequencies. 

\noindent $\bullet$ The expected flux density of the jet at low radio
frequencies -- e.g. 74 MHz -- is 17 mJy (if the spectrum breaks at 1.4 GHz)
and up to 74 mJy (break at or below 74 MHz). This level of emission will be
detectable by the proposed Long Wavelength Array (LWA) which will offer
arcsecond-resolution at MHz frequencies (Harris 2005). Current VLA 74 MHz
observations provide an integrated flux limit from GB~1508+5714 of $\sim$100
mJy (1$\sigma$) measured from the FITS map obtained by the VLSS survey
(Cohen et al. 2004). The integrated low-frequency radio emission may well be
dominated by the kpc-scale jet as radio core spectra are typically flat or
inverted ($\S$~\ref{radiospec}), and its emission is expected to be mostly
self-absorbed in the LWA-bands. If there is "contaminating"  emission from
the radio core and/or sub-arcsecond scale jet, astrometric uncertainties of
order 1'' or less in the LWA observations are then necessary to distinguish
emission from the different components. 

\noindent $\bullet$ The break in the electron spectrum constrains the peak
of the Compton component to a narrow range (formally between 5 and 160
keV).  The steep cm-wave spectrum we have measured implies that IC/CMB
emission from the highest-redshift jets may be minimal in the GLAST bands
($>$10 MeV). If these steep cm-wave spectra persist in other high-z jets,
GLAST\footnote{The main limiting factor of GLAST studies of kpc-scale jets
will be its inability to separate the expected gamma-ray emission from their 
small-scale jets with its course angular resolution.} studies of IC/CMB X-ray
jets may be restricted to lower-redshift targets with detected optical
synchrotron jets (e.g. Sambruna et al. 2004). 

\subsection{Fast-jets or sub-Equipartition Magnetic Field Strengths?} 

The one-zone synchrotron+IC/CMB model invokes relativistic motion on the
observed kpc-scales in order to preserve equipartition (i.e. avoid very
large total energies). This appears at odds with previous radio jet
asymmetry studies which set upper limits on the jet speed of order
$\Gamma\sim$3 (Bridle et al. 1994; Wardle \& Aaron 1997; Hardcastle et al.
1999). Although the bulk beaming factor necessary to explain the level of
X-ray emission in the GB~1508+5714 jet is not very large, other known
examples show that $\Gamma$'s up to $\sim$25 are necessary (e.g. Harris \&
Krawcynski 2002;  Kataoka \& Stawarz 2005).  One way to relax the speeds
invoked is to assume that kpc-scale quasar jets are not in equipartition
as discussed recently in Kataoka \& Stawarz (2005).  In this scenario, the
jets are characterized by sub-equipartition field and are particle
dominated.  This leads to quite unfavorably larger total energy
requirements and synchrotron models may provide a better solution (e.g.
Atoyan \& Dermer 2004). 


\subsection{X-ray Emission from Doppler-hidden beams?}

To maintain both equipartition and to accommodate radio studies which
inferred slower jet speeds than in the one-zone synchrotron+IC/CMB model,
one can invoke additional structure in the jet. Such
"two-fluid"  jet models have in fact been discussed for some time. Early
radio studies advocated "transverse" velocity structures with the observed
radio emission originating from a slower moving outer sheath which hides
emission from a faster moving spine (see summary by Bridle 1996). 

In powerful radio-galaxies and lobe-dominated quasars, we are
preferentially viewing the slower-moving outer radio-emitting jet layer
whose beaming cone is wide. In this case, the IC/CMB X-rays viewed by us
originate predominantly from the same observed radio emission and the
one-zone synchrotron+IC/CMB model is a good approximation -- Chandra/radio
observations of these jets may truly be telling us about their physical
conditions. 

In this picture, the X-rays "reveal" the
faster jet spine in the core-dominated quasars
which is responsible for most, if not all of the IC/CMB
emission (Celotti et al. 2001). With additional transverse velocity
structures come naturally other sources of high-energy emission such as
turbulent interactions between the spine-sheath layers (Stawarz \&
Ostrowski 2002).  In the IC/CMB scenario, the outer portion of the jet
producing most of the observed radio emission will certainly emit IC/CMB
radiation also, especially at high-redshift.  The key is not whether there
are IC/CMB X-rays, it is to consider also how much is being produced by
the hypothetical fast spine-layer (Jorstad \& Marscher 2004), and if its radio luminosity is 
only a fraction of the observed, how do we use the X-rays observations to tell us something
useful about its physical properties? 

In several kpc-scale jets in core-dominated quasars, the X-ray emission
terminates abruptly, while the radio emission continues on for 10's kpc or
more (e.g. 0827+243, Jorstad \& Marscher 2004; PKS~0605--085,
PKS~1510--089, and 1642+690, Sambruna et al. 2004). Do these mark the
decollimation points of the fast spines? One can indeed find other
core-dominated quasars where there is roughly 1-1 correspondence between
the radio and X-ray emissions throughout the jets (e.g. Sambruna et al. 
2004;  Marshall et al. 2005), and lobe-dominated quasars whose X-ray jets
do not persist to the terminal radio features (e.g. PKS~1136--135,
Sambruna et al. 2004). This may be a resolution dependent effect, as two
of the largest angular-size jets ($\sim$30'') show multi-wavelength
offsets in the terminal features (PKS~1127--145, Siemiginowska et al.
2002; PKS~1354+195, Sambruna et al. 2004) with the X-rays leading the
radio. 

How fast can the spine-beams be? Presumably from the observed X-ray
emission, one can deduce physical properties of the spines with the usual
machinery (section 2.2), as a function of fraction, $f$, of the detected
radio emission. As an analog to past studies of jet interaction with the
accretion disk photon field (e.g. Phinney 1987, Melia \& K{\"o}nigl 1989),
the implied beaming factors should not be so high as to avoid Compton drag
by the CMB photons, although this may be a way to explain the widely
observed declining X-ray to radio flux ratios observed in many jets,
which can be interpreted as jet-deceleration on the observed kpc-scales
(Georganopoulos \& Kazanas 2004). If also the beaming cones are made too
narrow ($\sim$1/$\Gamma$), one has to explain why we are finding X-ray
emission from the majority of prominent radio jets in core-dominated
quasars\footnote{See http://hea-www.harvard.edu/XJET/ for a list.}.

\section{Searching for the Highest-Redshift Radio Jets with the VLA}

Most Chandra studies of quasar jets have so far targeted {\it known}
arcsecond-scale radio jets (e.g. Sambruna et al. 2004, Marshall et al.
2005), and most are found at z$<$2 (Liu \& Zhang 2002). In fact, not many
z$>$2 radio jets (kpc-scale) are known. The most extensive radio imaging
studies of high-redshift quasars to date are VLA snapshot observations (few
min. scans) of quasars at redshifts of up to only $\sim$3 (Taylor et al.
1996; Barthel et al. 2000, and refs. therein). From these studies, only a
handful of arcsecond-scale jets are extended enough to be imaged with
Chandra. 

The majority of z$>$3 flat-spectrum radio quasars have only been discovered
over the last few years from large surveys (e.g.  Hook et al. 2002 and
references therein), and systematic radio imaging studies have yet been
carried out -- we are aiming to remedy this deficiency with a VLA
imaging survey. 

\begin{figure}
\includegraphics[width=60mm]{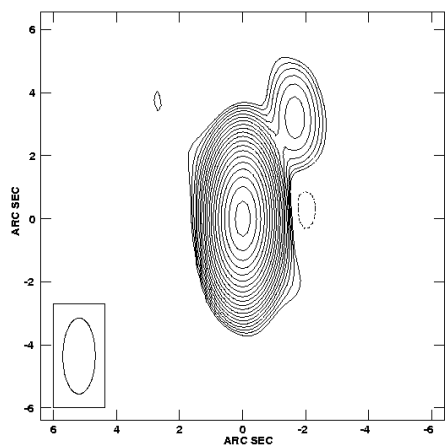}
\caption{VLA 1.4 GHz contour image of the z=3.78 quasar PKS~2000-330 showing an extended
radio feature to the northwest of the quasar. The
contours begin at 0.5 mJy/bm and increase by increments of $\sqrt{2}$. The beam is 2.41 
$\times$
1.03'' at a PA of --0.2 deg.
} \label{f2} \end{figure}

\subsection{Target Selection}

We used NED\footnote{This research has made use of the NASA/IPAC
Extragalactic Database which is operated by the Jet Propulsion Laboratory,
California Institute of Technology, under contract with the NASA.} to
help compile a sample to search for extended radio jets with the VLA with
the aim of followup Chandra imaging. 
For an initial sample, we selected objects with redshifts of 3.4 and greater, and radio 
fluxes \simgt 100 mJy at 1.4 and/or 5
GHz. We
excluded known radio galaxies and steep spectrum sources (e.g. van Breugel
et al. 1999; De Breuck et al. 2001) and focused on the flat-spectrum
ones (a first order proxy for high beaming) to facilitate a comparison
with the known z$\simlt$2 X-ray jets in core-dominated quasars.
Our goal was to compile the best
candidates to be imaged with the VLA in order to determine how common
large-scale jets are at high-redshift, rather than to create a complete
sample. Imaging the fainter objects and extending to lower redshifts
(z$\sim$3--3.4) are obvious extensions to our work and this is planned. 

We then performed a literature and VLA archive search of the 43
sources in our final sample (z=3.4--4.8): 9 have known radio structures
(jets, gravitational lenses), or have sufficient archival VLA data
(Table~1). Our final targets were drawn from the remaining 34
objects which did not have suitable VLA data available. Six objects are at
declinations of \simlt $-30^{\circ}$, so may explain why so little
archival VLA data are available for them (see below).

\begin{figure}
\includegraphics[width=55mm,angle=-90]{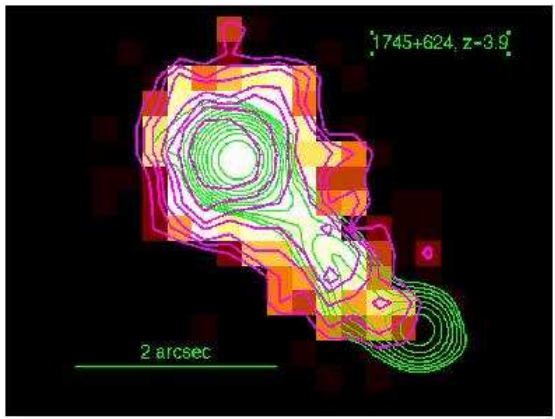}
\caption{Chandra color (and magenta contours) and VLA 5 GHz (green 
contours) images of the jet in 1745+624 (4C~62.29) at z=3.9. The X-ray data are
taken from the Chandra archive (PI: P. Strub; OBSID 4158). Details can be found in
Cheung et al. (in preparation).
} \label{f3} \end{figure}

Over nine observing runs in 2004 Oct and Dec, we observed 26 targets with
the VLA-A configuration at 1.4 and 5 GHz (resolution 1 and 0.4 arcsec).
Since the main goal of this survey was to find jets, we only obtained
short snapshots at 5 GHz to estimate spectra relative to the deeper 1.4 GHz
images.  Archival VLA data exist for a handful of other sources
satisfying our criteria and these have been analyzed also. This led to the
discovery of a 3.6'' long radio jet in one quasar, PKS~2000--330
(Figure~2). Archival Chandra data were available for the z=3.9 quasar 1745+624, which 
revealed an X-ray counterpart (Figure~3) to its known radio jet (Becker et al. 1992).

The details of this survey will be presented elsewhere. In summary, extended
radio features are detected in $\sim$50$\%$ of the sample, although only
a handful are extended enough in the z \simgt 3.4 sample to be imaged with Chandra's 
$\sim$arcsecond resolution. The most distant object we found with a jet is the z=4.72 
quasar GB~1428+4217 (Fabian et al. 1997). 
Proposed X-ray observations will allow us to determine
the relative importance of IC/CMB versus synchrotron X-ray emissions at the 
highest-redshifts, leading to an important test of expectations between the X-ray 
emission models.

\section{Summary}

Jets are common features of radio-loud AGN, at radio and X-ray 
wavelengths, at least locally (z$\simlt$2). Our work suggests 
that this may be true also at high-redshift (out to z=4.7) at radio wavelengths. 
Proposed Chandra observations will tell us if this is true also at X-rays. 
Extending studies to these high-redshifts should help us to distinguish 
between the competing synchrotron and IC models to determine if there is a redshift 
dependence in the X-ray jet emission. 

\begin{table*}[t] 
\begin{center}
\caption{High-redshift (z$>$3.4) Quasars with (publicly) Known Radio 
Structures} 
\begin{tabular}{|c|c|l|}
\hline\hline
 Name       & z   & Description [References]\\
\hline
  0201+113 &   3.61  & known 4'' radio jet \cite{sta90}\\
  1239+376 &   3.818 & known $\sim$6'' radio jet \cite{tay96} \\
 1351--018 &   3.707 & VLBI-scale jet \cite{fre97} \\
  1422+231 &   3.62  & gravitational lens \cite{pat92} \\
  1508+572 &   4.301 & known $\sim$2.5'' X-ray/radio jet \cite[][This work]{sie03,yua03,che04} \\
 1630--003 &   3.424 & gravitational lens \cite{win02} \\
  1745+624 &   3.889 & VLBI jet \cite{tay96}; $\sim$2.5'' X-ray/radio jet \cite[][This work]{bec92} \\
 2000--330 &   3.773 & VLBI jet \cite{fom00}; new $\sim$4'' radio jet (This work)\\ 
  2215+020 &   3.572 & VLBI jet; diffuse 7'' radio extension \cite{lob01} with possible X-ray emission \cite{sch02b} \\ 
\hline                                      
\end{tabular}
\end{center}
\end{table*} 

\bigskip
\begin{acknowledgements}

C.~C.~C. is grateful to {\L}ukasz Stawarz and Alan Bridle for insightful 
discussions, and Dan Harris for useful comments on the manuscript. 
The National Radio Astronomy Observatory which is
operated by Associated Universities, Inc. under a cooperative agreement
with the National Science Foundation (NSF). 
Radio astronomy at Brandeis University is supported by the NSF through grant AST
00-98608. 

\end{acknowledgements}

\end{document}